\documentclass[11pt]{article}

\usepackage[utf8]{inputenc}
\usepackage[T1]{fontenc}
\usepackage[english]{babel}

\usepackage[margin=1in]{geometry}
\usepackage[protrusion=true,expansion=false]{microtype}
\usepackage{array}
\usepackage{calc}
\usepackage{booktabs}
\usepackage{longtable}
\usepackage{ragged2e}

\usepackage{amsmath}
\usepackage{textcomp}

\usepackage[round,authoryear]{natbib}
\usepackage{xcolor}
\definecolor{linknavy}{RGB}{0,51,102}
\usepackage[breaklinks=true]{hyperref}
\usepackage{url}
\hypersetup{
  colorlinks=true,
  urlcolor=linknavy,   
  citecolor=black,     
  linkcolor=black,     
}
\newcommand{\doi}[1]{\href{https://doi.org/#1}{doi:#1}}

\usepackage{needspace}
\let\oldsection\section
\renewcommand{\section}{\needspace{8\baselineskip}\oldsection}
\let\oldsubsection\subsection
\renewcommand{\subsection}{\needspace{3\baselineskip}\oldsubsection}

\title{\textbf{Open Veins of AI Auditing}\\[0.3em]
  \large How AI Scrutiny Lags Deployment in the Global South}
\author{Gemma Galdon Clavell \and Alexandra Magaard}
\date{An Eticas Foundation publication.\\
  Working draft for arXiv submission (as a pre-print to future publications).}

\begin{document}

\maketitle

\begin{abstract}
Artificial intelligence systems are being deployed across the Global South at a pace that now matches, and in some sectors exceeds, deployment in select countries in the Global North. In both regions the infrastructure of AI governance has not accompanied that deployment, but the situation, as we will show, is much worse in the Global South. Drawing on a decade of AI audit practice, including the only fully published second-party algorithmic audit of a deployed AI system in the region (Robot Laura, Brazil), a completed but unreleased second-party audit of a national child-welfare risk prediction system in Latin America, a second, also unreleased, audit of a national public employment matching algorithm in the region, a set of rapid pandemic-response assessments conducted in 2020, thirteen Responsible AI Assessments conducted across Sub-Saharan Africa and Asia Pacific, and a regional landscape analysis for Latin America and the Caribbean, this paper documents patterns found in the sparse field of AI evaluation in the Global South and explains why it so rarely occurs. We count less than twenty published second- and third-party algorithmic audits and assessments of deployed systems across Africa, Latin America, the Caribbean, South Asia, and Southeast Asia over the past decade, against hundreds of documented public-sector algorithms and multibillion-dollar national government investments in AI. Our research finds four cross-cutting patterns from the AI evaluations that we have conducted in the Global South: proxy targets that substitute predictability for validity, AI performance claims that collapse under prevalence analysis, populations scored by models that never saw them in training, and patterns of structural bias in model outputs that continue even after removing protected attributes (e.g. demographic data) from training data. We then identify five lessons for evaluation, regulation, and funding that emerge when AI evaluation moves outside the regulatory, linguistic, and data conditions of the Global North. We argue that this gap between the Global North and the Global South is not, at root, a capacity problem but a funding problem: capacity follows funded demand, and across the Global South's rapidly growing AI deployment, no actor is currently required, or funded, to hold a deployed system to account. Finally, we identify the actor best placed to close this gap: the small number of development and philanthropic funders that stand behind most consequential AI in the region, and whose funding conditions can require independent evaluation where no regulator yet does. We close with clear recommendations to that end for funders, governments, and the design of evaluation itself, aimed at improving the state of AI evaluation, and arguably the state of AI as a whole, in the Global South.
\end{abstract}

\vspace{0.5em}

\vspace{1em}

\section{Introduction}

Over the past five years, the deployment of artificial intelligence in the Global South has moved from smaller government-funded pilot projects to deep integration in public infrastructure. Researchers have documented 234 algorithms operating in the public sector in Latin America alone, concentrated in social protection, health, education, justice, and public security \citep{smart2024}. The African Union adopted a Continental AI Strategy covering all 55 member states in July 2024 \citep{africanunion2024}. India's national AI mission was approved with a budget of roughly 1.25 billion US dollars \citep{govindia2024}; Brazil's national AI plan commits approximately 4.25 billion US dollars through 2028 \citep{govbrazil2024}. Industry projections place AI's 2030 contribution at up to 2.9 trillion US dollars for Africa \citep{gsma2024}, with PwC's more conservative estimate of 1.2 trillion covering Africa alongside other developing markets \citep{pwc2017}.

At the same time, there is an emerging market of AI auditors, evaluators and overseers. Yet that emerging market, supposed to verify whether these systems work, and for whom, is overwhelmingly concentrated in institutions and firms headquartered in the Global North. Across all of the regions in the Global South, namely Africa, Latin America, the Caribbean, South Asia, and Southeast Asia, Eticas has counted only a very small number of publicly documented second and third-party audits of deployed AI systems over the past decade (Section 2). The United Kingdom alone has a denser concentration of AI assurance providers than the publicly documented audit activity across the Global South. The infrastructure behind ISO/IEC 42001, the international AI management standard, tells the same story: the recognized accreditation bodies driving it are almost all Northern, as are the certification bodies they accredit and the organizations certified so far. Adoption across the Global South is nascent at best; India's national accreditation body, for instance, has only recently begun adapting its framework to accredit AI management-system certification, while documented uptake across Sub-Saharan Africa, Latin America, and most of South and Southeast Asia remains minimal. This is in any case a low bar, since ISO/IEC 42001 certifies an organization's governance processes and documentation, not the behavior of any deployed system.

These regions lack even a minimum layer of assurance, let alone the capacity to evaluate how deployed AI actually performs on the ground. The professionals who perform algorithmic audits are concentrated in a handful of Global North countries, evaluating systems and platforms whose harms are studied almost exclusively on Western populations \citep{urman2025}. This is problematic in many ways, not least because these AI systems are often deployed on vulnerable populations in contexts where they have limited defense and advocacy capacity to begin with, and where governments have less power to demand that providers take matters of fairness, accountability and transparency seriously.

Over the past decade, Eticas is, to our knowledge, the only actor that has consistently conducted second- and third-party evaluations of planned and deployed AI systems across all of these regions. A small body of published work by others exists, and we document it in Section 2, but even what exists is clustered in a handful of Latin American countries and in India while leaving Sub-Saharan Africa, Southeast Asia, and the Caribbean almost entirely uncovered. It is also worth mentioning that while we would have liked to be able to develop a consistent capacity to audit in the Global South, we have had to rely on scattered funding and our own limited resources. As we will show below, this has meant that, due to funding conditions, we have not always been able to apply a consistent method to our evaluations. Nonetheless, these years of work have allowed us to describe dynamics, barriers and possibilities that must be part of any take on the impact of AI on the Global South, and we hope that this paper contributes to calling attention to the need to build capacity and infrastructure for AI evaluations beyond the North.

We have built this article on six different engagements:

\begin{itemize}
\item
  A second-party algorithmic audit of a deployed AI system in Brazil. This remains, to our knowledge, the only fully published audit of this kind in the region. The system, ``Robot Laura,'' is a clinical deterioration prediction system that by 2021 had processed more than 8.6 million patient visits across 40 clinical and hospital centers. The audit was commissioned by the Inter-American Development Bank and published in 2021 \citep{eticas2021};
\item
  A second-party audit, conducted between 2019 and 2020, of a national child-welfare risk prediction system in Latin America that scores 3.9 million children using 280 variables drawn from seven administrative sources, including education records. The completed report was delivered in March 2020 but never released; we report its findings here in anonymized form for the first time (Section 3.1);
\item
  Three rapid pandemic-response engagements for the Inter-American Development Bank in 2020, counted here as a single cluster: (i) a fixed-scope ethics screening of a portfolio of AI systems seeking funding to scale during the COVID-19 emergency, later adopted by the funder as a standing internal screening tool; (ii) the design of a regional governance package for contact-tracing technologies across Latin America; and (iii) technical support to a national government's pandemic digital response. This work built on the foundational ethics framework for AI systems in development programs that we had co-developed for the same institution in 2018-2019;
\item
  A second-party audit of a national public employment matching algorithm in a Latin American country (2020-2021), examining bias and fairness, explainability, and accountability in how job seekers were matched to vacancies. The audit drew on the platform's full registration and matching records for 2018 to 2020, covering a database of more than 25,000 registered job seekers;
\item
  Thirteen Responsible AI Assessments of systems deployed across Sub-Saharan Africa and Asia Pacific in partnership with the German International Development Organization, GIZ's FAIR Forward initiative, spanning agriculture, climate and disaster response, energy access, public services, and language technology (LLMs);
\item
  A study for the United Nations Development Programme (UNDP) taking stock of the AI market across Latin America and the Caribbean, the impact of biased AI on the region's population, and the region-specific conditions that shape how AI is built and deployed there, published as a background paper for its 2025 Regional Human Development Report.
\end{itemize}

This body of work constitutes a meaningful share of the world's empirical record on Global South AI evaluation. We reference our own work not to generalize from our own practices, but simply out of necessity: together with the small number of published evaluations by others documented in Section 2, our research accounts for the largest share of what exists. It was this realization that prompted this paper: we write to share what our practice has taught us, and to make it easier for others to fill a gap that no single organization can close.

In the next sections, we make three central claims. First, that the AI evaluation gap in the Global South is quantifiable and growing rather than closing, and that the governance wave of 2024 to 2026 (AI safety institutes, summit declarations, readiness assessments) will not close it: those institutions are calibrated to frontier models and country-level diagnostics rather than deployed systems, a limitation they share with their Northern counterparts, with the difference that the Global North retains other layers of scrutiny beneath them while in most of the Global South this miscalibrated layer is the only one that exists (Section 2). Second, that evaluation under Global South conditions is both feasible and productive: every system we examined showed concrete room for improvement, and every engagement produced the metrics, mitigation measures, and design recommendations needed to realize it, from recalibration paths and subgroup performance disclosure to data protection fixes. On our record, evaluation is not a brake on AI adoption but an input to systems that work better for the populations they serve, and the practice surfaces five methodological issues that evaluation frameworks built for Northern conditions do not address (Sections 3 and 4). Third, that the gap persists because there are no actors committed to closing it: evaluation capacity is thin, in the Global South as everywhere, but capacity follows demand, and demand exists only where someone with leverage requires evaluation and funds it. Today, no major funder of consequential AI in the Global South makes independent evaluation a condition of its funds, no actor with leverage over deployed systems is obliged to scrutinize them, and publication rather than regulation is what converts an audit into accountability. The actors best placed to change this, and the addressees of our recommendations, are the funders whose funds stand behind many of the systems (Sections 5 and 6).

Our central argument, underpinning all our sections, is that AI oversight in the Global South is missing not because it is unwanted or impossible but because no one prioritizes it: current international and national government approaches to AI development either overlook oversight entirely, defer to voluntary governance frameworks, or are too generic to reach the consequential systems already deployed, and the funders who could change this by making independent evaluation a condition of their funds have not yet done so.

A note on scope and terminology. We use "Global South" as shorthand for Africa, Latin America and the Caribbean, South Asia, and Southeast Asia, acknowledging the term's limitations. We use "audit" in a strict sense: a structured, independent assessment or evaluation of a specific deployed or near-deployment system that examines training data, model behavior, or impacts, and produces documented findings. Please note that throughout the text we distinguish audits from investigative journalism, advocacy, policy mapping, and country-level AI readiness diagnostics. These are all valuable (and more abundant) tools that complement the oversight picture. We apply the same strictness to our own work: of the six engagements on which this paper draws, three are audits under this definition (one published, two unreleased), one is a portfolio of structured assessments that we deliberately do not call audits, one is a set of rapid pandemic-response assessments, and one is a landscape analysis.

\section{The State of AI in the Global South: Deployment Without Evaluation}

\subsection{Hundreds of AI systems already deployed}

As of summer 2026, there are records of 234 known and documented algorithms within Latin America's public sector alone. Sixty percent of those algorithms have been designed in the past four years, and there is little to no public information on their deployment status (are they in active use?), ownership (which government institution owns and operates them?), or performance, even to the researchers who identified them \citep{smart2024}. No similar picture exists for any other region in the Global South.

Because so little is known about the AI systems in use, government investment offers a second indicator of the scale of deployment. A continental AI strategy spans all 55 African Union member states, and the 2025 Africa Declaration on Artificial Intelligence announced a 60 billion US dollar continental AI fund aimed at infrastructure, talent, and AI enterprises \citep{africanunion2025}; India's national AI mission is funded at roughly 1.25 billion US dollars; and Brazil's national AI plan commits approximately 4.25 billion US dollars through 2028, of which less than half of one percent is allocated to AI oversight and governance.

\subsection{Less than twenty published audits and assessments}

When one considers only second and third-party independent audits, arguably the most robust form of AI evaluation, we could only find a handful tackling deployed AI systems in the Global South over the past decade.

In 2021, Eticas and the Inter-American Development Bank published an audit of \emph{Robot Laura}, a clinical deterioration prediction system operating in Brazilian hospitals. The audit identified gaps in Brazilian patients' data protection and how AI outputs fit into existing infrastructure and decision-making processes \citep{eticas2021}. In 2019 and 2020, Eticas audited a national child-welfare risk prediction model in another country of the region. In this case, the completed report has never been released (Section 3.1 presents its findings in anonymized form); the audit found that the prediction model functioned in practice as a detector of poverty, frequently scoring children it had never seen in training. In 2020 and 2021, Eticas also audited a national public employment matching algorithm in a Latin American country; that report, too, was never released, and we share our finding, in anonymized form, here for the first time. As of now, five years later, Robot Laura remains, to our knowledge, the only fully published second-party audit of a deployed AI system in Latin America; the child-welfare and employment audits, though completed, were never released. A fourth case is the Applied Artificial Intelligence Laboratory at the University of Buenos Aires' third-party audit of an adolescent pregnancy prediction system deployed in Salta, Argentina. Their audit reverse-engineered the system's published methodology and identified contamination in the system's training data and bias within its database \citep{pedace2022}. A fifth case comes from litigation: in 2022, court-ordered technical examinations of Buenos Aires' fugitive facial recognition system, obtained through a constitutional challenge brought by the civil-society organizations ODIA and CELS, documented that the system had been used to run queries on more than 15,000 people who were not on the fugitive list it was legally restricted to, and the system was declared unconstitutional \citep{cels2022}. Beyond audits in the strict sense, Chile's GobLab at Universidad Adolfo Ibáñez has since 2021 conducted collaborative bias and transparency assessments of more than a dozen deployed public-sector models, published as case summaries rather than full audit reports \citep{goblab2025}, and Fundación Karisma published a documentary analysis of Colombia's Sisbén IV social classification algorithm, conducted without access to the system's data or code \citep{lopez2020}. We count these as assessments and analyses, not as independent audits.

In South Asia, we have found three academic audits, albeit with noteworthy limitations. All three of those audits were based in India (and none in any other country within the heterogeneous South Asian region) and all three audits addressed commercial application programming interfaces (APIs) rather than deployed public AI systems. All three adapted the Gender Shades methodology developed in \citet{buolamwini2018} to Indian faces and voices. They comprise an audit of four face-processing APIs \citep{jain2021}, an adversarial audit of facial recognition services \citep{jaiswal2022}, and an extension to automatic speech recognition, where audits of commercial ASR systems for Indian users show how label quality itself can distort audit results \citep{mishra2021}. All three documented the same pattern: performance that holds on Western benchmarks degrades sharply when the same services are pointed at Indian faces and voices. Beyond commercial APIs, two bodies of work address deployed public systems in India: LibTech India's transaction-level assessments of Aadhaar-based wage payments under the national rural employment guarantee, which analyzed tens of millions of transactions and documented payment delays and rejections disaggregated by caste \citep{bheemarasetti2023}, and Amnesty International's attempted algorithmic audit of Telangana's Samagra Vedika entity-resolution system, which was denied access to the system and published its methodology and human rights findings instead \citep{amnesty2024}; that a well-resourced international organization could not complete an audit for lack of access says as much about the accountability environment as the count itself. In Sub-Saharan Africa, Southeast Asia, and the Caribbean, our team did not find any fully published second or third-party audit of a deployed AI system. The published record is thus not merely thin; it is uneven, and for entire regions it is empty. The closest body of work is our own portfolio of thirteen Responsible AI Assessments conducted with GIZ FAIR Forward across Sub-Saharan Africa and Asia Pacific (Section 3.3), published as an open methodology in 2024 \citep{giz2024} and updated in 2026. We count these within the evaluation record, but not as audits: they were light-touch assessments, their methodology adapted to GIZ's programmatic needs rather than our full audit protocol.

The strict count, then, remains vanishingly small relative to deployment: fewer than twenty cases in ten years, and by the strictest definition roughly ten to fifteen, depending on inclusion criteria, across regions that are home to the majority of the world's population \citep{un2024}.

\subsection{An Emerging AI Evaluation Industry Based Solely in the Global North}

A new industry of AI auditors is growing in the Global North. The United Kingdom government's 2024 study of the market identified 524 firms supplying AI assurance goods and services. This included 84 specialized AI assurance companies, generating approximately 1.01 billion pounds in gross value added and employing over 12,500 people \citep{dsit2024}. Projections for this market vary widely with definitions: the UK government projects its domestic AI assurance market alone to exceed 6.5 billion pounds by 2035 \citep{dsit2024}, while one independent market study projects the global market to grow from 1.63 billion US dollars in 2023 to 276 billion by 2030 \citep{aiat2024}.

Within this new industry, its own layer of professional infrastructure follows the same pattern: the founding leadership of the International Association of Algorithmic Auditors (IAAA) sits entirely in the United States, the United Kingdom, the European Union, and Canada;\footnote{Disclosure: the first author of this paper is the current President of the IAAA \url{https://www.iaaa-algorithmicauditors.org/}} and a field scan of the algorithmic auditing ecosystem found that even among self-identified practitioners, only 37 percent had ever participated in an audit, with the interviewed field leaders drawn entirely from US and UK institutions \citep{costanzachock2022}. The research base mirrors the practice base: a systematic review of platform audit studies found 84 percent used Western samples and 65 percent of authors held Western institutional affiliations \citep{urman2025}.

\subsection{No structural mandate to oversee AI deployment}

Since 2024, there has been a wave of AI governance institution-building, which could have focused on closing the evaluation gap. However, that is not what the new AI governance institutions and policies have been designed for. To illustrate this point, for example: the International Network of AI Safety Institutes includes eleven governing bodies and nations. Only two of those, Kenya and Singapore, are from the Global South. Furthermore, these newly founded institutes have an average budget of 10 million dollars and staffs of a few dozen people, and they are institutionally designed to monitor only the safety of frontier AI models and their capabilities in bio and cyber risk, loss of control, and election interference. What none of these institutions is designed to do is look at systems already in use: their mandates include no post-deployment monitoring, no measurement of impacts on the populations that deployed systems score and serve, and no accountability for the AI harms that are documented and occurring now rather than anticipated. The institutions built to govern AI's future have no tools for AI's present.

This structural gap is demonstrated in the International Network of AI Safety Institutes' output. Their first report, the International AI Safety Report \citep{bengio2025}, focused its substantive agenda on risks from frontier models. There were no specific research directives for any kind of AI evaluation in the Global South, and this dynamic has persisted in their two recent ``key updates''.

The implication is uncomfortable but clear: the most consequential AI accountability gap in the Global South sits outside the mandate of every institution created in the current governance wave, and will therefore not be closed by it. The AI systems that allocate welfare assistance, access to health, score children's risk, and help citizens navigate benefits, education opportunities and agricultural credit are essentially exempt from scrutiny.

\section{What the Few Audits from the Global South Reveal}

This section presents four evidence streams from our own practice over the past decade. It covers the main types of AI systems deployed (child-welfare risk scoring, clinical prediction, agricultural and public-service AI, and the broader regional landscape), the main types of AI evaluation (published audit, unreleased audit, structured assessment portfolio, landscape analysis), and the main outcomes from those AI evaluations (publication, non-publication, methodology release, policy uptake).

\newpage

\begin{longtable}[]{@{}
  >{\raggedright\arraybackslash}p{(\columnwidth - 8\tabcolsep) * \real{0.1524}}
  >{\raggedright\arraybackslash}p{(\columnwidth - 8\tabcolsep) * \real{0.2236}}
  >{\raggedright\arraybackslash}p{(\columnwidth - 8\tabcolsep) * \real{0.1931}}
  >{\raggedright\arraybackslash}p{(\columnwidth - 8\tabcolsep) * \real{0.1931}}
  >{\raggedright\arraybackslash}p{(\columnwidth - 8\tabcolsep) * \real{0.2378}}@{}}
\toprule\noalign{}
\begin{minipage}[b]{\linewidth}\raggedright
\textbf{Stream}
\end{minipage} & \begin{minipage}[b]{\linewidth}\raggedright
\textbf{System and region}
\end{minipage} & \begin{minipage}[b]{\linewidth}\raggedright
\textbf{Evaluation mode}
\end{minipage} & \begin{minipage}[b]{\linewidth}\raggedright
\textbf{Publication status}
\end{minipage} & \begin{minipage}[b]{\linewidth}\raggedright
\textbf{What it showed}
\end{minipage} \\
\midrule\noalign{}
\endhead
\bottomrule\noalign{}
\endlastfoot
3.1 Child-welfare risk model & National childhood alert system scoring 3.9 million children, Southern Cone & Second-party a posteriori bias audit (2019-2020) & Delivered March 2020; never released; findings reported here in anonymized form & Proxy substitution, misleading headline metric, out-of-scope scoring, bias surviving removal of protected variables \\
3.2 Robot Laura & Clinical deterioration prediction, 8.6 million patient visits, Brazil & Third-party audit under the IDB's fAIr LAC initiative (2021) & Published in full & Deliberate over-prediction differing by sex and age; data protection gaps; recommendations delivered to deployer \\
3.3 RAIA portfolio & Thirteen AI systems across Sub-Saharan Africa and Asia Pacific & Structured light-touch assessments (2022-2026) & Open methodology published (2024; updated edition forthcoming 2026) & Feasibility at low cost; tractable early-stage fixes; goodwill with nothing to reward it \\
3.4 Landscape analysis & AI, bias, and evaluation across Latin America and the Caribbean & Regional study for UNDP (2025) & Published as background paper to the 2025 Regional Human Development Report & 234 documented public-sector systems and near-zero evaluation; region-specific risk structure \\
\end{longtable}

\textbf{Table 1.} The four evidence streams of Section 3.

\subsection{Auditing a national child-welfare risk model in the Southern Cone}

Between 2019 and 2020, we conducted a bias audit of the predictive risk model underlying a national childhood alert system in a Latin American country. Our audit was developed for the ministry responsible for social and family policy and prompted by the international institution that had funded the system in a rare case of funder-led inspection that allowed us broad access to data. In line with our confidentiality obligations, we identify neither the country nor the commissioning institution. Our findings are reported at the level of technical detail we can share. The completed audit report was delivered in March 2020, but never published by the commissioning institutions. We summarize its principal findings here because they illustrate, with unusual precision, what deployed systems look like when someone finally inspects them; the non-publication of the report illustrates the rest of this paper's argument.

The national child-welfare risk model we audited scores children for their risk, aiming to have an objective picture of children that need specific support because their rights are at risk. It is used, monthly, to generate prioritization lists for local childhood offices in pilot municipalities, with alerts suggested for children in the 95th percentile of estimated risk. The model integrates 280 variables (describing the child, mother, father, siblings, household, and environment) from seven administrative sources, including the census, education records, criminal records, and the child protection agency's own registries, over a dataset of 3.9 million children. We analyzed all the data that was made available to us under a confidentiality agreement, and we organized our audit's findings into four areas.

First, we found that the target the model predicts is not the harm the system is described as preventing. Put differently, the system's outputs are communicated as a child's risk of vulnerability, but what the model actually estimates is the child's proximity to a specific administrative intervention. Within our audit, our team tested five candidate proxy variables during development, and found that the proxy variable selected (entry into a child protection program involving separation from the family) was chosen because it was the variable the model could best predict, not because it was the proxy closest to the concept of rights violation. Our audit documented the gap: many children suffer severe rights violations without ever entering a separation program, because of capacity limits, judicial decisions, or because their cases never reach the agency at all. In practice, this means the system does not direct caseworkers toward the children at greatest risk; it directs them toward the children who most resemble past administrative caseloads, while the children the state has never managed to see, often the most vulnerable, remain invisible to the tool that is supposed to find them. Our audit recommended that this distinction be explicitly and proactively stated to every user of the system, and that the stated goals of any AI system are a true reflection of the data choices made within it.

Second, we found that the model's headline performance metric was misleading. The event the model predicts is rare, occurring for roughly 3 in every 1,000 children. The reported metric, an area under the ROC curve of 0.94, measures how well the model ranks children across all possible alert thresholds, and with an event this rare it is inflated by the model correctly ranking the overwhelming majority of children who were never at risk. It says nothing about performance at the threshold the system actually uses to trigger alerts. At that operating point, the model's detection rate was 73 percent: roughly one in four children who were in fact separated had been scored as not at risk. In practice, the two numbers point in opposite directions: the 0.94 is an argument for trusting the system, the 73 percent is a warning against relying on it alone, and a caseworker who deprioritizes an unflagged child because the model is "94 percent accurate" is making exactly the error the headline metric invites. We recommended that user-facing communication center on the detection rate, not the ROC, so that caseworkers understand what the model misses.

Third, the differential impact analysis, conducted with the Aequitas toolkit on the test partition, found systematic disparities. Predicted risk rose steeply with age (a disparity ratio of 0.77 for ages 0 to 3 against 1.8 for ages 14 to 17). Children with national citizenship were predicted to be at high risk at 1.62 times the rate of foreign children, in a context where migrant children are likely underrepresented in the data because many lack a national identification number; the group most plausibly at elevated risk was the group the model was least equipped to see. Children from indigenous communities received higher predicted risk than non-indigenous children (1.0 against 0.78) even though indigenous status had been deliberately excluded from the model; the exclusion did not prevent the disparity, it only made it impossible to monitor through the model's own variables, a textbook demonstration that when the system developer removes protected attributes from the dataset, all that is achieved is more confusion or opacity rather than fairness. The presence of information about the mother doubled the probability of a high-risk prediction (2.2), while the pattern for father ran in the opposite direction (0.35); the model uses 40 variables describing mothers and 15 describing fathers, encoding the assumption that maternal circumstances are the source of risk and paternal circumstances barely relevant. This discrepancy has obvious gendered implications that our team recommended be rebalanced. Children with a family member who had passed through the child protection agency were almost nine times more likely to receive a high-risk prediction, a pattern that may capture genuine risk transmission but also institutionalizes a feedback loop in which once a child comes in contact with the agency, so does their family, whether justifiably or not. Finally, a child's predicted risk score increased as their parental education and municipal income fell. This confirmed Eticas' original concern, raised publicly when the system was announced, that the model would function in practice as a poverty detector.

Fourth, the audit identified multiple structural defects in the system's scope and robustness. Variables with near-zero prevalence in the training data (e.g. maternal substance abuse at 0.3 percent, maternal mental disability at effectively zero) carried high coefficients, meaning that the model was too sensitive to signals that are too rare for it to estimate reliably. We recommended that these variables be altogether removed from the statistical model and handled with explicit rules for reserved intervention quotas. Most strikingly, children aged 16 to 17 and children lacking parental information were excluded from the model's training and testing. However, the model scores these children anyway: a population of children is assessed every month by a model that never saw anyone like them during training.

The audit produced a full set of mitigation recommendations: explicit proxy variable disclosure, detection-rate-centered communication, subgroup effectiveness analyses, capture of indigenous and migrant status for evaluation purposes only, ordinal treatment of education variables, territorial transfer testing, and rules-based handling of rare high-stakes signals. Whether any were adopted, we cannot say, because the report's non-publication forecloses exactly the public verification that audits exist to enable. As of now, the Eticas team assumes the system continues to operate, scoring children in real time. We report the findings here, six years later, and note that a regime in which the publication of audit findings about a system scoring millions of children is discretionary is not a regime with audits in any meaningful sense.

\subsection{Auditing a clinical prediction AI tool in Brazil}

Robot Laura is an AI-driven system designed to predict patient deterioration in hospitals across Brazil, providing real-time alerts to healthcare teams. By 2021, it had processed more than 8.6 million patient visits across 40 clinical and hospital centers in five years. Eticas audited Robot Laura's system as part of the Inter-American Development Bank's (IDB) fAIr LAC initiative, and the full audit and findings are publicly available on its website \citep{eticas2021}.

Eticas' audit combined a literature review, interviews with developers and clinical personnel, an analysis of personal data administration, and a bias and fairness audit for differential impact by sex, age, and other intersectional categories. The quantitative analysis used the same Aequitas toolkit as the audit described in Section 3.1, working from 2,874 hospital records supplied by the developer from a southern Brazilian hospital where the system operates, including data on patients' sex, age, hospital sector, the model's outcome probability and threshold, the binary prediction, and the observed outcome. The system itself classifies a patient's risk of deteriorating from their last five vital-sign collections over a 24-hour window. This includes indicators, like oxygen saturation, respiratory rate, blood glucose level, and blood pressure, and the threshold for when each indicator triggers an alert is negotiated locally with each hospital's medical team.

Our findings were striking and differed between age groups and by sex. The model's training target is in-hospital mortality, and it deliberately over-predicts it at roughly double the observed rate for both sexes (31.2 percent predicted against 17.4 percent observed for men; 26.2 against 12.3 for women), a calibration that suppresses false negatives (which remain at 5 to 6 percent) at the cost of precision. Once our team started analyzing by specific age group and sex combined, the differences between results became even more pronounced. For example, the system most often underestimates deterioration risk precisely for younger women, and its calibration curve degrades in the middle of the score range, where the gap between the sexes is widest. Against a target band of 0.8 to 1.25 on the disparity ratios, positive predictive value for patients aged 18 to 39 was 67.9 percent for men but 39.3 percent for women (1.35 against 0.78, both outside the band), and the false-negative disparity exceeded the threshold for women aged 18 to 39 (1.43) and 40 to 59 (1.32).

Our recommendations included warning hospital administrators and staff that the system underestimates risk for women aged 18 to 39; suggesting recalibration around the cut-point and re-test the effect on subgroup precision and false-negative rates; monitor or exclude groups too sparse to model reliably, as the audit itself did with the handful of minors in the dataset; and publish model documentation (objectives, data, methodology, and performance and error margins by group) to staff and patients.

The audit also identified concrete paths to better protect patient data and to integrate the system's outputs into clinical workflows in ways that preserved professional judgment, and the published report strived to make the findings useful for both the deployer and the wider regional conversation on health AI.

Finally, the audit is itself arguably more important than any of our specific findings, because the Robot Laura audit demonstrates that it is possible and replicable to conduct a second-party audit of a deployed AI system in a Global South context, if a funder deems this work important.

\subsection{Thirteen AI assessments across Africa and Asia Pacific}

Between 2022 and 2026, in partnership with GIZ's \emph{FAIR Forward initiative}, we conducted thirteen Responsible AI Assessments of AI systems developed and deployed across Sub-Saharan Africa and Asia Pacific, and consolidated the methodology into an open, publicly available guide, the Responsible AI Assessment Guide (RAIA), first published in 2024 \citep{giz2024}. The guide has since been revised and expanded into an updated edition, forthcoming at the time of writing, and it is that updated edition that informs this paper \citep{giz2026}. The assessed systems map the actual texture of AI deployment in these regions, which differs sharply from the consumer and productivity applications that dominate Northern evaluation research. Specifically, while evaluations in the North often focus on consumer applications and social media platforms or general-purpose chatbots, in Sub-Saharan Africa and Asia Pacific we audited systems related to:

\begin{itemize}
\item
  machine-learning detection and classification of buildings from imagery to predict rural energy demand and site off-grid electrification in Uganda;
\item
  an NLP chatbot making entrepreneurship policy documents searchable and usable in Ghana;
\item
  an AI-powered micro-climate forecasting and advisory chatbot, and a conversational chatbot for discovering digitized government services, both in Kenya;
\item
  computer-vision identification of cashew crop disease from field images in Ghana;
\item
  a retrieval-augmented generation platform delivering agricultural advice to smallholder farmers and extension officers in Kenya and India;
\item
  AI solutions for coastal communities in Indonesia, consisting of citizen-science data collection designed to feed locally developed AI models released as digital public goods;;
\item
  AI classification of forests by carbon-storage capacity to prioritize protection in India;
\item
  a retrieval-augmented chatbot summarizing public-sector audit reports to make them accessible and actionable in Uganda;
\item
  AI-supported coffee yield estimation from guided field photographs in Uganda;
\item
  a Kenyan-languages corpus built as training infrastructure for local-language AI;
\item
  machine-learning landslide detection and early warning in Rwanda;
\item
  AI crop-type mapping from earth observation data in Telangana, India.
\end{itemize}

Our thirteen assessments measure these systems against five of the risk categories of the Eticas AI Risk Taxonomy \citep{eticas2026taxonomy}: bias and fairness, privacy and confidentiality, reliability, governance, and security and misuse. These categories operationalize the UNESCO Recommendation on the Ethics of AI \citep{unesco2021}, with distinct guidance for automated decision-making systems and for generative systems built on large language models. Each assessment follows a defined workflow from pre-survey and scoping through analysis and a structured deep dive to a risk and mitigation report with follow-up, examining risks at the pre-processing, in-processing, and post-processing stages of the system lifecycle.

What the assessments surfaced is revealing. The projects arrived willing: teams across the portfolio shared documentation, code repositories, training data, and models, and sat through scoping calls and structured deep dives alongside local domain experts and the Eticas team. This voluntary exposure of developers to outside scrutiny was, again, facilitated by the funding agency supporting all of these initiatives. Because most systems were assessed early in the lifecycle, at the pre-processing stage, the recommendations that emerged were tractable: the Guide's worked example for the Ugandan coffee-yield system runs to rebalancing gender representation in the training data and anonymizing farmer records before any public release, fixes that cost little when a system is young and become expensive after deployment hardens design choices. The pattern repeats across the portfolio: for Rwanda's landslide early-warning system, the assessment recommended integrating demographic and construction data so that risk maps track how exposure actually evolves rather than reproducing unrepresentative historical records; for Telangana's crop-mapping system, it recommended hardening the classification model against low-quality field images, standardizing and incentivizing farmer survey inputs, and making crop-area definitions and sampling transparent before the maps become the basis for agricultural planning. The portfolio's generative systems add a further layer: for the retrieval-augmented platform delivering agricultural advice to farmers in Kenya and Bihar, India, the assessment examined content grounding, source selection, and prompt manipulation, and paid particular attention to a question no model benchmark can answer: when multiple organizations contribute the data, the models, and the services, who is responsible for accuracy, updates, and harm mitigation once the advice reaches a farmer? This assessment also illustrates what we expect to become the generative-era pattern: the team had invested seriously in grounding, with outputs validated against curated question sets and retrieval-quality metrics and human review for sensitive topics, while the characteristic blind spots sat elsewhere, with no structured defenses against prompt injection, no AI-specific incident response, and governance documentation lagging well behind the system's technical sophistication.

The uncomfortable finding is structural: there was goodwill but nothing to reward it. A team that implements every recommendation earns no certification, no procurement advantage, no regulatory credit, no access to better funding, and whether mitigation happens depends, as the Guide itself concedes, on each team's commitment and resources. The region's problem is not only that bad systems go unexamined; it is that good-faith builders have no support structure that makes their diligence count.

Three features of this portfolio bear on the argument of this paper. First, the assessments demonstrate feasibility: structured, technically grounded evaluation can be conducted with development teams in low-resource settings, across sectors from agriculture to disaster management, within the documentation, data, and infrastructure constraints those settings impose. The methodology was co-created and tested with partners across both regions precisely so that it reflects real-world constraints, including varying levels of digital infrastructure, institutional capacity, regulatory maturity, and data availability. Second, while these were light-touch assessments adapted to GIZ's programmatic needs rather than full audits, they still managed to identify material, actionable risks in every system examined, and help development teams increase their risk awareness and take concrete action. Third, the methodology is published, open, and reusable \citep{giz2024,giz2026}, and each assessment makes the next one cheaper: the templates, question sets, and worked examples already exist, so a new team pays only the cost of applying the method, not of building it..

\subsection{Zooming out: AI, bias, and evaluation in Latin America and the Caribbean}

Our analysis for the United Nations Development Programme of gender bias in AI across Latin America and the Caribbean \citep{galdon2025} provides the wide-angle view that the system-level work cannot. Its central empirical observation is epistemic: of the 234 public-sector algorithms documented in the region, information on who operates them, whether they remain deployed, how they work, and whether they have ever been evaluated is mostly unavailable, not only to the public but to the researchers who compiled the inventories. The deployment record is itself unaudited; the count of systems is a lower bound, and the count of evaluated systems is, with the exceptions documented in this paper, approximately zero.

The landscape analysis also surfaces the region-specific risk structure that Northern evaluation frameworks do not capture: language asymmetry, in which systems nominally operating in Spanish or Portuguese reason over training distributions dominated by English and Chinese ("even when the AI speaks Spanish, it thinks in English," in the formulation we documented from regional researchers); digital dependence, in which the systems making consequential decisions about the region's populations are developed, hosted, and governed elsewhere; the gendered texture of deployment, from the feminization of public-facing chatbots (Violetta, Sara, Sofía, Laura) to documented cases such as a woman falsely accused in Brazil after a facial recognition misidentification during Carnaval; and an economic asymmetry in which AI is projected to contribute 5.4 percent of regional GDP by 2030 against 14.5 percent in North America, ensuring that the region absorbs AI's decisions faster than it captures AI's value. These are not exotic edge cases; they are the operating conditions of AI in the region, and Section 4 develops their methodological consequences.

\subsection{The pattern across the four streams}

Read together, the four evidence streams support three conclusions. The first is that evaluation finds things: in every case where we examined a system, we surfaced material issues, from proxy substitution and misleading headline metrics to out-of-scope scoring, bias that survived the removal of protected variables, and gaps in data protection, that the deploying institutions had not identified, and in several cases could not have identified, because the relevant analyses had never been run. In most cases, the findings were linked to issues of equity, vulnerability, digital inclusion, climate resilience, environmental sustainability and accountability itself.

The second is that this work is feasible under Global South conditions, and feasible cheaply: the published audit, the unpublished audits, the thirteen-system assessment portfolio, and the regional landscape study were all completed within the data, language, and institutional constraints this paper describes, using methods that are public and reusable, and all of them were small engagements, run by small teams on modest budgets. Cost, like capacity, follows practice: these engagements are inexpensive for us because several hundred audits and assessments over the past decade have taught us where to look and how to measure, and the published methodology now encodes that experience for anyone else to use. The absence of evaluation, in other words, is explained by neither technical impossibility nor cost.

The third conclusion is the one that structures the rest of the paper: everything documented in this section happened at a funder's discretion. Every engagement in our record exists because a development bank, a development agency, or a UN program chose, once, to pay for scrutiny; none exists because of a legal obligation or funding requirement. This discretion also sets the ceiling on what the work can achieve: in our case, the audit a funder chose to publish informed its deployer and a regional policy program, while the audit left to the commissioning institution's discretion was suppressed, despite being the more alarming of the two. A field that exists only where someone happens to ask for it is not accountability infrastructure; it is a series of fortunate accidents. Converting the accident into a system is a funding decision, and it is the decision to which the rest of this paper is addressed.

\section{Five Structural Challenges for the Future of AI Auditing in the Global South}

The methods of algorithmic evaluation are often designed with ideal background conditions in mind: administrative data that exists and covers the population; benchmarks tailored to a system's working language; a regulator with the mandate and capacity to eventually read the report and act on it; recourse mechanisms for the people affected; and a developer, a deployer, and an affected population within reach of the same jurisdiction. These ideal conditions are not true in the North, which may be a reason why so much talk about AI auditing and evaluations struggles to become practice, as regulations and standards hit the messy reality of complex supply chains, missing data and a trade that requires a technical capacity most oversight bodies still lack. These go a long way in explaining why audits and accountability have failed to yet generalize in the North.

While even in the North these conditions hold only partially, in the Global South they mostly cannot be assumed at all: the data is thinner, the benchmarks speak the wrong languages, the regulator may not yet exist, and the supply chain crosses several jurisdictions before it reaches the person affected. The difference is one of degree, but degree is what decides whether a method transfers. Drawing on the practice documented in Section 3, this section distills five lessons for anyone who evaluates, funds, or regulates AI in the Global South. Each is invisible from the center and obvious from the field, and together they are an argument about who should do this work and who must pay for it: evaluation capacity must be built in the regions where the systems run, by practitioners for whom these conditions are the baseline rather than the exception.

\begin{longtable}[]{@{}
  >{\raggedright\arraybackslash}p{(\columnwidth - 4\tabcolsep) * \real{0.2337}}
  >{\raggedright\arraybackslash}p{(\columnwidth - 4\tabcolsep) * \real{0.3354}}
  >{\raggedright\arraybackslash}p{(\columnwidth - 4\tabcolsep) * \real{0.4309}}@{}}
\toprule\noalign{}
\begin{minipage}[b]{\linewidth}\raggedright
\textbf{Lesson}
\end{minipage} & \begin{minipage}[b]{\linewidth}\raggedright
\textbf{The failure it prevents}
\end{minipage} & \begin{minipage}[b]{\linewidth}\raggedright
\textbf{What it requires}
\end{minipage} \\
\midrule\noalign{}
\endhead
\bottomrule\noalign{}
\endlastfoot
4.1 Evaluate impacts, not models & Systems that pass benchmarks and fail their populations & Outcome data, community ground truth, independent modeling of registry blind spots \\
4.2 Regulate the right things & Rules too broad to bite, or specific in the wrong ways & Disaggregated outcome metrics, lifecycle checkpoints, local-language test assets \\
4.3 Supply chains and dependencies & Accountability without an addressee; single-vendor lock-in & Obligations that travel by contract; open, model-agnostic assets \\
4.4 State capacity as the binding constraint & Technology-first systems that outrun the institutions they depend on & Evaluation in context; AI evaluation joined to impact evaluation as one continuum \\
4.5 Quantification with consequences & Findings that change nothing & Publication by default; measurement enforced by those who fund the systems \\
\end{longtable}

\textbf{Table 2.} Five lessons for AI evaluation in the Global South (Section 4).

\subsection{Evaluate impacts, not models}

The current governance wave evaluates models: their capabilities, their benchmark scores, their behavior under test conditions. But a model is not yet a system, and a system is not yet an impact. Frontier-model evaluation sits at the maximum possible distance from the point where AI meets real, complex users. And yet, most evaluation energy in the North seems to be targeted at understanding these high level, highly-technical, remote and IP-protected models. In our work here and elsewhere \citep{eticasfoundation2026}, we have shown how an impact approach facilitates community involvement, forces evaluators to tackle AI system behavior in ``the wild'' and not a lab or synthetic context, situates policy in a space that is known and relatively comfortable, and focuses enforcement efforts at a point that can send powerful messages upstream: if impact will be evaluated, mitigating harmful impacts better start at the top of the funnel and continue through the supply chain.

This approach also questions the idea of the AI black box, which is, in our experience, largely a myth. Our community-led audits from the last ten years have repeatedly evaluated deployed systems from an adversarial, third-party position, using outcome data, crowdsourced evidence, and experimental testing \citep{eticasfoundation2026}, and our Global South engagements confirm that what evaluation mostly requires is not access to model weights but attention to outcomes.

Outcomes are also where the data is most valuable. The most informative measure of a deployed system is rarely its model outputs; it is what happened to the people scored, flagged, or advised, today and in the long term. And yet currently, impact data does not even exist. Developers have no incentive to collect it and deployers often have no technical capacity or infrastructure to access and use it.

Even when evaluators and auditors have access to model data, impact data is crucially relevant. Where administrative registries fail to see whole populations, the model's blind spot reproduces the registry's, as with the migrant children of Section 3.1, and evaluation must independently model the missing population rather than inherit the gap. Community impact data and documented user experience fill precisely the gaps that datasets leave open.

In resource and capacity-constrained contexts like many in the Global South, the shift from models to impacts makes real accountability feasible, affordable and policy-effective, as it measures AI system against the world they actually operate in. Methodologies that cannot incorporate this knowledge default to what they do have: benchmarks and test sets built elsewhere, for other populations.

\subsection{Regulate the right things: the Northern record as a cautionary tale}

The Global South is routinely urged to catch up with Northern AI regulation. Our practice suggests a different reading of that record: Northern Ai regulation is a catalogue of calibration errors that the Global South still has the chance to avoid. Northern instruments have tended to fail in one of two directions. Some are too broad: principles-based frameworks that invoke fairness, transparency, and accountability but define no test, no metric, and no checkpoint, and therefore produce commitments rather than findings. Others are specific in the wrong ways. New York City's Local Law 144, the first mandated bias-audit regime for hiring algorithms, requires an annual, model-output audit against aggregate impact ratios: fewer than 5 percent of covered employers published an audit, and those that did disclosed that their results complied with acceptable thresholds. But systems that pass aggregate tests can still fail specific groups at specific decision points \citep{wright2024,galdon2026employment}. Rules that are specific in the wrong way measure what is easy to measure, not what uncovers harm..

The technical twin of this error is the benchmark. Test sets exist overwhelmingly in English and a handful of high-resource languages; for most African languages, and for the regional varieties of Spanish, Portuguese, and Asian languages in which deployed systems actually operate, equivalent benchmarks are thin or absent. A system can therefore pass every test available to its auditor and still fail the population in front of it, in the language, register, and dialect that population actually uses, and the gap compounds for generative systems, whose outputs must be judged for fluency, cultural appropriateness, and factual accuracy exactly where evaluation resources are weakest. The right specificity, on our record, looks like the opposite of both errors: disaggregated outcome metrics rather than headline scores (Section 3.1's detection-rate lesson), lifecycle checkpoints rather than annual certificates, publication requirements rather than filing requirements, and local-language test assets treated as shared infrastructure. Regions that have not yet legislated can adopt this calibration directly.

\subsection{Supply chains, dependencies and the procurement chokepoint}

The separation of where an AI system is built from where it acts on people is a structural feature of the global AI supply chain, one that even the European Union, with considerable regulatory weight, has struggled to address. What differs in the Global South is leverage. The child-welfare model of Section 3.1 was developed with an academic center on another continent; the systems in our assessment portfolio run on cloud infrastructure and foundation models governed from other hemispheres; the consumer systems documented in Latin America and the Caribbean are overwhelmingly developed, hosted, and updated abroad. A jurisdiction that hosts none of the compute, holds none of the documentation, and represents a small share of the provider's market may have no effective addressee for accountability demands at all: when the model card sits in one jurisdiction, the training pipeline in a second, and the affected children in a third, each actor can truthfully say the obligation belongs to someone else.

Philanthropy may now be adding a new layer to this dependency structure. In May 2026, Anthropic and Gates Foundation announced a 200 million US dollar partnership committing grant funding, model usage credits, and technical support to global health, education, and economic mobility programs across low- and middle-income countries \citep{anthropic2026}. While the partnership has the ambition this paper calls for, and will fund real services and public datasets for underserved populations at a moment when open alternatives remain weaker, it also concentrates the stack: a generation of health, education, and agricultural deployments will be built on a single commercial model, evaluated against benchmarks developed by that model's own vendor, and exposed to an uncertain future once subsidized access fades. Two design choices could convert partnerships of this kind from dependency into infrastructure: independent evaluation of the systems funded, and investment in open, model-agnostic evaluation assets that outlive any single provider.

\subsection{State capacity as the binding constraint}

The hardest constraint we encounter in the Global South is not technical but institutional. Many of the systems we assessed arrived technology-first: built to demonstrate what AI can do, with interoperability, maintenance, and institutional fit treated as somebody else's later problem. An early-warning model for landslides is only as valuable as the state's capacity to move a warning from a dashboard to a village, and to protect and rescue once it arrives; a benefits chatbot is only as valuable as the services it points to; a risk score is only as valuable as the caseworkers, budgets, and protocols that act on it. In several of our engagements the most consequential findings concerned not the model but the delivery chain around it: data infrastructures that could not sustain the system, institutions without the mandate or staffing to respond to its outputs, and no plan for either.

This is why evaluation must happen in context. An assessment that scores the model and ignores the institution measures the system's potential, not its effect, and a funder relying on it is buying a demonstration, not a capability. It is also why AI evaluation and impact evaluation belong on one continuum: development actors have spent decades building rigorous impact-evaluation practice, with baselines, counterfactuals, and outcome tracking. AI evaluation should enter that toolkit as its newest instrument rather than growing beside it as a parallel bureaucracy. The registry gaps of Section 4.1 are a symptom of the same constraint: what an institution cannot see, no model can fairly score, and no evaluation can fully verify. Evaluating in context means measuring the system and the capacity it depends on as one object.

\subsection{What is not quantified cannot be changed}

Our final lesson is the oldest one. Modern philanthropy and modern advocacy were built on measurement. Booth's poverty maps of London and Rowntree's study of York turned deprivation that polite society treated as anecdote into data that could be contested; the Russell Sage Foundation's Pittsburgh Survey of 1907-1908, philanthropy-funded and unapologetically empirical, documented industrial labor conditions with a rigor that helped professionalize social work, fed the campaigns that won workmen's compensation, and opened reform politics to people who were not elites. What was counted could be changed; what was not counted, for political purposes, did not exist. Quantification was not a technical exercise but the creation of a political space, and it was philanthropy that led the effort.

The modern record shows the same mechanism wherever evaluation has been enforced. When medical journal editors made trial registration a condition of publication in 2005, registrations at ClinicalTrials.gov tripled within two years, and among large publicly funded cardiovascular trials the share reporting positive outcomes fell from 57 percent before mandatory registration to 8 percent after, a direct measure of the distortion that unmeasured practice had been hiding \citep{kaplan2015}. Public-access mandates reached over 80 percent compliance once enforcement was attached to grant payments. The Equator Principles adopted by banks in 2003 requiring borrowers to conduct environmental and social impact assessments, built new professions, community recourse and awareness across the very regions discussed in this paper.

Publication is the hinge of the mechanism. As our own record demonstrates, the impact of an evaluation is governed not by its quality but by whether its findings enter the public record: the audit a funder chose to publish informed its deployer and a regional policy program, while the audit left to a commissioning institution's discretion informed no one, despite being the more alarming of the two. Where regulation is absent, the publication decision is the accountability decision, and leaving it to the audited party converts evaluation into a private consulting product. For evaluations to become accountability, they must produce public traces.

The question this record leaves is simple: who, in the Global South's AI ecosystem, can impose measurement with consequences? Regulators could, eventually; Section 2 showed why they have not yet. As shown above, large funders can fund the practices and incentives that will change current power relations around AI. Nowhere is that leverage greater than in the Global South, where a small number of institutions stand behind most consequential AI developments, and where a funding policy focused on impact, evaluation and transparency can reach further than any statute.

\section{Future Outlook: What Philanthropy Can Build}

The preceding sections have argued that the evaluation gap is a stable equilibrium, held in place because no actor with power over deployed systems is also obliged to scrutinize them. Of the actors surveyed in this paper, one is positioned to break this. Most consequential AI in the Global South is funded, directly or through governments, by a small number of institutions, and to date they have asked grantees to commit to responsibility rather than required them to evaluate their systems. Major philanthropic programs ask applicants to show awareness of responsible-AI considerations rather than to evaluate what they build. The World Bank's Environmental and Social Framework, the instrument that reshaped safeguard practice across a generation of infrastructure lending, contains no reference to AI at all. Even the European Union, which positions itself as the home of human-centric and trustworthy AI, does not carry that ambition down into the conditions attached to the funds it spends: an analysis of two decades of EU research funding found that its commitment to "human-centric, sustainable, secure, inclusive and trustworthy artificial intelligence" does not travel into funding practice, with less than a third of AI-related calls making any reference to trustworthiness, privacy or ethics and responsibility \citep{galdon2023}. Annual dedicated funding for AI in development contexts, an estimated 350 to 600 million US dollars, is concentrated in roughly a dozen institutions, and to our knowledge not one of them yet makes independent evaluation a binding condition of funding. Every deployed AI system in the Global South backed by a major development or philanthropic actor today is funded without one.

The most recent wave of philanthropic commitment illustrates the pattern rather than breaking it. Since 2024, several large initiatives have promised to steer AI toward the public interest: Current AI, launched at the 2025 Paris AI Action Summit, opened with 400 million US dollars and a stated target of 2.5 billion over five years; Humanity AI, a coalition of ten foundations, pledged 500 million over five years; and the Patrick J. McGovern Foundation committed 73.5 million to "AI for humanity" across 144 organizations. These are serious sums, and the intent behind them is real. Yet the grants disbursed to date flow to policy research, advocacy, journalism, literacy and community engagement, and, to our knowledge, none of the three makes independent evaluation of deployed systems a condition for the funds it moves. The largest and newest public-interest AI commitments thus reproduce the very gap they were created to close, funding awareness of the problem rather than verification of the systems \citep{currentai2025,humanityai2025,mcgovern2024}.

The consequences of this continued pattern are felt in the maturity of alternatives to Big Tech everywhere, but especially in the Global South. Funders have a first move advantage that can change the game just as they did in the late nineteenth and early twentieth centuries. Governments act on evidence, and in most of the Global South the evidence base that would justify AI oversight, the documented failures, the quantified harms, the demonstrated value of evaluation, does not yet exist at scale, because no one is required to produce it. A concerted effort to scale audits and evaluations, where every project creates a use case, every evaluation adds to a shared record of data, benchmarks, and reusable infrastructure, every published finding creates urgency, and every quantified harm creates the political space in which regulators, procurement officers, and advocates can act, is the theory of change that this paper proposes.

\subsection{Five commitments}

Our asks are five, and they fit in a paragraph. We address them to funders because funders can move first, but the same five conditions work identically in any instrument that directs funds toward an AI system: a philanthropic grant, a development loan, a procurement contract, a national budget line. Make independent evaluation a condition of funding, proportionate to each system's power over people, continuous from design through deployment, and budgeted inside the deployment budget rather than beside it. Point evaluation at impacts, not models, with communities and local experts convened as ground truth. Integrate AI risk into the machinery funders already run, both the vetting funnel through which every grantee and vendor enters and the impact-evaluation cycles through which every program is judged. Secure publication contractually, before findings exist to be suppressed. And spend on open, model-agnostic assets rather than vendor-tied ones, so that each engagement builds the field. Five clauses in a grant agreement; what follows from them is larger than it sounds. One condition runs beneath all five: the capacity to evaluate must be built and held in the regions where the systems operate, so that conditioning funds transfers ownership southward rather than re-importing Northern terms under a new name.

\subsection{What a decade of evaluation infrastructure would create}

A decade of that practice would leave the Global South holding assets that no AI ecosystem, North or South, holds today. None of them is exotic; each has a precedent in a field that once faced the same void and built its way out.

\textbf{Data for longitudinal studies}: a continuous, comparable record of what deployed systems actually do to people's health, learning, work, and welfare, kept up as providers and models change, so that harms, benefits and impacts, individual and collective, can be tracked over years rather than glimpsed in a single evaluation and then lost. Development has built exactly this kind of asset before. The Demographic and Health Surveys, funded by USAID since 1984, created standardized, comparable data across some ninety countries and turned public health in much of the world from guesswork into something measurable; the Framingham Heart Study followed one town's residents for decades and gave medicine the very idea of a "risk factor"; climate science has the Mauna Loa carbon record, running since 1958. AI has no equivalent, and no single actor in the ecosystem with a plan to develop it yet.

\textbf{Methods at experimental grade}: evaluation reliable enough to answer the question that matters to anyone paying for a system, namely whether it actually helped compared with not using it. Development economics made precisely this leap within a single generation. The randomized trial moved out of medicine and into anti-poverty programs through groups such as MIT's J-PAL, an approach recognized with the 2019 Nobel in economics, and a claim that a cash transfer or a tutoring program works is now expected to be tested against a comparison group, not simply asserted. AI-for-development today sits roughly where that field sat in the 1990s: full of pilots, thin on evidence that any of them beats the alternative, precisely at a time when the cost of randomized trials could be significantly reduced thanks to AI.

\textbf{Models, benchmarks and systems that make sense locally:} benchmarks that can tell whether a system works in the languages and contexts people actually live in, models built for those languages and contexts in the first place and systems that bring capacities together to solve actual problems with AI. These are already being assembled from the ground up, often by communities of volunteers (Masakhane, Mozilla's Common Voice, AI4Bharat) and even at the regional level (Lelapa AI's InkubaLM for African languages, AI Singapore's SEA-LION for Southeast Asia, and the CENIA-led Latam-GPT, built openly by institutions across fifteen Latin American countries). These models cost a fraction of frontier systems to train and run and can be governed where they are used. A clinic chatbot does not need a model that writes sonnets in forty languages; it needs one that understands how patients in its region describe pain and when to hand over to a human. Small, context-specific models are infrastructure in exactly this paper's sense: assets whose value accrues locally and that loosen the dependency documented in Section 4.3. All of these efforts remain underfunded and atomized; an effort to being resources together and facilitate their maintenance and distribution would convert scattered goodwill into shared infrastructure that the North, which also lacks these resources, may end up borrowing.

\textbf{Systems built for response capacity}: evaluation that rewards systems designed to work with the institution around them, not just in the lab. The principle is old in humanitarian practice. FEWS NET, a famine early-warning network used since 1985, was built on the hard lesson that a hunger forecast is worthless unless food and money can actually move in response to it. A landslide alert with no one to carry the warning to the village, or a benefits chatbot that points to services which do not exist, is the AI version of the same mistake. Rewarding the whole chain rather than the dashboard is what turns a demonstration into a capability.

\textbf{A local industry that builds and holds accountable}: a home-grown profession made of the people who build these systems and the people who check them, with the training, the credentials, and the standing to refuse bad work. Entire professions have been conjured this way before. US securities law in the 1930s created the modern financial-audit industry out of almost nothing; the World Bank and IFC safeguard rules, and the Equator Principles cited earlier in this paper, built an environmental and social assessment profession across the developing world where none had existed. Each began as a condition attached to funds and ended as a career path. What this paper argues for is the same move in AI, with one difference: this time the senior tier sits in the region, not only in Northern head offices.

\textbf{Open infrastructure}, in two senses. Open \emph{to scrutiny}: public registers of which algorithms governments actually use, so that citizens, journalists, and researchers can find them in the first place. Early versions already exist; Amsterdam and Helsinki publish municipal algorithm registers, the United Kingdom runs an Algorithmic Transparency Recording Standard, and Chile's GobLab, cited earlier, maintains a public repository of state algorithms. Open \emph{to build with}: shared models and datasets that let local teams create without renting their foundations from a single vendor, in the spirit of the Human Genome Project, whose 1996 decision to release its data publicly seeded an entire open biotechnology economy. One commons serving both scrutiny and creation, so that accountability and innovation stop competing for the same budget line and start reinforcing each other.

An ecosystem holding these assets, and more, escapes the love/hate narrative that has captured the AI debate, because it can answer the only question that narrative avoids: what does this system, here, actually do? The initiatives now forming, with their hundreds of millions committed to people-centered, public-interest AI, will write their ten-year retrospectives either way; whether those documents describe systems bought or a field built is the choice their funding clauses are making now.

\subsection{Levers beyond funders}

Funders are the fastest-moving lever, not the only one. Governments hold a powerful instrument in procurement: demanding relevant tests and metrics, robust and independent evaluations, concrete guarantees around safety, proper incident reporting, and clear roles, responsibilities, and limits to liability turns each major acquisition into a demand signal for local evaluation capacity, and survives political cycles better than standalone AI statutes. Regional accreditation bodies can close the certification vacuum documented in Section 2, so the emerging assurance market does not simply import Northern certifiers. Country-level instruments such as the UNESCO readiness methodology can add a deployed-system audit component to their follow-up cycles, converting readiness diagnostics into accountability pipelines. And South-South practitioner networks can pool the methods, benchmarks, and language assets that no single country's market yet sustains. Each reinforces the others; none requires waiting for the rest.

\section{Conclusion}

AI systems are already in the Global South, and their harms have been documented: proxy targets standing in for children's welfare, models scoring populations they never saw, structural bias that persists even after potentially biased variables are removed. These harms are real and evidence-based, documented by the small number of evaluations permitted to share their findings, in regions where the consequences fall on those with the least recourse.

The methods to oversee these systems exist, and the practitioners to run them exist in small numbers that will grow exactly as fast as there is work for them to do. What is missing is the requirement, and the institutions best placed to create it are, ironically, the ones whose funding stands behind the systems themselves. AI in the Global South is one funding decision away from the transformation that scrutiny brought to medicine and finance: not because oversight is cheap or easy, but because the actor who can order it already holds the pen.

The title of this paper alludes to Eduardo Galeano's book, \emph{Open Veins of Latin America}. Galeano described how Latin America's veins had been opened so that others could profit from what flowed out. AI runs through the same channels: the data, the decisions, and the value move north, and what returns is systems no one can check or hold accountable. None of that is inevitable. The ideas and tools to close the veins already exist. They are being developed in the trenches of AI impact and evaluation. But they are small and atomized, with no single actor having the resources or the incentives to build field-wide infrastructure. Funders need only decide that a system worth deploying is a system worth examining. Whether the next decade is remembered as more of the same, or as the moment the Global South began to keep what it builds and account for what it runs, is not a technical question but a choice.

\section*{About this work}

This paper is a publication of the \textbf{Eticas Foundation}, the nonprofit arm of Eticas. Eticas operates a deliberately hybrid, open-core model: Eticas.ai, the commercial practice, conducts algorithmic audits and AI assessments for public and private clients, and that revenue-funded work generates the methods, practitioner experience, and empirical record. Eticas Foundation, the 501(c)(3) nonprofit organization that publishes this paper, turns that record into public goods: open methodologies, published research, community-led audits, and policy engagement aimed at building the field of accountable AI rather than any single organization's position within it. As we continue to cater to under-served actors, we have recently added Lebanon, Palestine and Türkiye to our list of use cases and hope to publish results soon.

\nocite{eticas2020}

\bibliography{references}

\end{document}